\documentclass[prd,aps,showpacs,showkeys]{revtex4-1}
\usepackage{bm,amsfonts,latexsym,amsmath,amssymb,amsbsy,graphicx}

\newcommand{\bb}{\begin{eqnarray}}
\newcommand{\ee}{\end{eqnarray}}
\newcommand{\ba}{\begin{align}}
\newcommand{\ea}{\end{align}}

\begin{document}

\title{\bf  On the creation of charged massless fermion pair by a photon  in crossed electromagnetic field}
\author{V.R. Khalilov} \email{khalilov@phys.msu.ru}
\affiliation{Faculty of Physics, M.V. Lomonosov Moscow State University, 119991,
Moscow, Russia}

\begin{abstract}
Creation of charged massless fermion pair by a photon in  external constant crossed electromagnetic field is considered. For this we use the expression of elastic scattering amplitude (EAS) of photon  in the one-loop approximation of massive quantum electrodynamics obtained earlier and calculate its massless limit. We assume that  the imaginary part of EAS of photon describes the total probability of  charged massless fermion pair creation in external electromagnetic field. Photon emission by a charged massless fermion is also studied in constant crossed electromagnetic field. We obtain the total probability of photon emission calculating elastic scattering amplitude of charged massive fermion in the  electromagnetic field in the massless limit.
\end{abstract}



\maketitle


\section{Introduction}

Electrodynamics effects with taking part of charged massless particles attract attention in last time.  The problem of electromagnetic radiation from  massless charges was considered in  \cite{1a,2a, 3l,4k,jko} in the framework of classical electrodynamics. The problem of photon emission by charged massless scalar particle in an external magnetic field was solved in \cite{gal} in massless scalar quantum electrodynamics.  Gal' tsov \cite{gal} proved that  the electromagnetic radiation of such a particle in external magnetic field is essentially quantum effect and must occur in the form of emission of hard photons with energy of the order of the particle energy. It was also shown in  \cite{gal} that
the (strictly) massless scalar quantum electrodynamics and  the zero-mass limit of the massive theory yield the same results for the total probability, the total radiation power and the spectral distribution of radiation.

It was shown in \cite{machet,mach2} that the one-loop self-mass of an electron of mass $m$ propagating in a graphene-like medium in a constant external magnetic field do not vanish at $m\to 0$.
In \cite{khgen,kh19} the  elastic scattering amplitude of planar  charged massless fermion in an  external constant  homogeneous magnetic field was obtained in the one-loop  approximation of the  2+1 dimensional quantum electrodynamics;
the total probability of photon emission by charged massless fermion as well as the one-loop  massless fermion self-energy were calculated.

Great interest to effects of quantum electrodynamics with taking part of charged massless fermions in external electromagnetic (especially Coulomb) fields  is related to problems of  graphene (see, for instance, \cite{7,8,9}).
It will be noted that important effect - the vacuum instability in the so-called  supercritical Coulomb potential - is supposed to occur in graphene with charged impurities. The instability of quantum electrodynamics vacuum  in supercritical Coulomb potential  of a hypothetical atomic nucleus with the charge  $Ze>Z_{cr}e\sim 170e$ has been studied for a long time and it is very important physical but highly academic problem  \cite{004,005,006,007,008}. It has been understood that this phenomenon is related to electron-positron pair creation from a vacuum.

In graphene, because the corresponding ``effective fine structure constant" is large ($\sim 1$) \cite{7,10,13}, a cluster of charged impurities  can produce  supercritical Coulomb potential which opens the real possibility of testing
the vacuum instability \cite{9}.  Supercritical vacuum instability manifests in  the creation of  quasi-stationary states with negative energies that are directly associated with the  positron creation in supercritical Coulomb potential in  quantum electrodynamics \cite{14}. The electron-hole pair creation (holes in graphene play the role of positrons (see, for example \cite{vp11a,as11b,as112}) is likely to be now revealed in graphene  \cite{gr0,gr1,gr2}.  The electron-hole pair production in graphene is the condensed matter analog of electron-positron pair production  due to the polarization and instability of the quantum electrodynamics vacuum in the supercritical Coulomb potential  \cite{004,005,006,007,008}.  Vacuum polarization of graphene  with a Coulomb impurity was addressed  in  works \cite{7,9,vp11a,as11b,as112,bsaso,kn11c,12,vmpvnk,gscr1,13a,yn1}.
Quasi-stationary states in supercritical Coulomb potential were studied in  \cite{khllee,khcr1,khcr2}. It was shown in \cite{khcr2} that the imaginary part of "energy" of quasi-stationary state is the doubled probability of the creation of  charged massless fermion pair by supercritical Coulomb potential.

It is of interest to consider one more effect of quantum electrodynamics: the creation of charged massless fermion pair by a photon in external electromagnetic field.
Here we study the problem of photon propagation in an external constant crossed electromagnetic field.
We show that hypothetical charged massless fermion pair can be created by a photon in  electromagnetic field.
We use EAS of photon obtained  in the one-loop approximation of massive quantum electrodynamics and calculate its massless limit. We assume that  the imaginary part of EAS of photon is related to the total probability of charged massless fermion pair creation. Photon emission by a charged massless fermion is also studied in the crossed electromagnetic field. We obtain the total probability of photon emission calculating elastic scattering amplitude of charged massive fermion in  crossed electromagnetic field in the massless limit.

We shall adopt the units where $c=\hbar=1$.

\section{Elastic scattering amplitude of photon in a crossed electromagnetic field}

The polarization operator (PO)  in a constant crossed electromagnetic field in the coordinate representation (in $e^2$ order) is determined by
\bb
\Pi^{\mu\nu}(x', x'';A)=-ie^2{\rm tr}[\gamma^{\mu}S^c(x',x''; A)\gamma^{\nu}S^c(x'',x'; A)],
\label{po0}
\ee
where $S^c(x',x''; A)$ is the causal Green function of the Dirac equation,$x\equiv x^{\mu}=x^0, x^1, x^2, x^3$, $A\equiv A^{\mu}=a^{\mu}n\cdot x$ is a  vector potential and $n^{\mu}$ is unit wave vector of crossed (plane-wave) electromagnetic field.
In the proper-time representation the Green function $S^c(x',x'', A)$ of
the Dirac equation for a fermion of mass $m$ and charge $e$ can be written as
 \bb
S^c(x',x''; A)=-\frac{1}{4\pi^2}\int\limits_{0}^{\infty}\frac{ds}{s^2}\exp\left[-\frac{ix^2}{4s}
-is[m^2+(eFx)^2/12] -iea\cdot x n\cdot x_+/2\right]\times \nonumber\\
\left[m+\frac{\gamma\cdot x}{2s}+\frac{e}{2}ms\sigma F-\frac{e^2}{3}s\gamma FFx +i\gamma^5\gamma F^*x\right],
\label{maggreen}
\ee
where $s$ is the ``proper time", $x\equiv x^{\rho}=x^{'\rho}- x^{''\rho}$, $F\equiv F^{\mu\nu}$ is a tensor,
$F^*\equiv F^{*\mu \nu}$ is a dual tensor of crossed electromagnetic field and $x_+\equiv x_+^{\mu}=x^{'\mu}+ x^{''\mu}$, $\gamma^{\mu}$ is Dirac's matrices, $\gamma^5=i\gamma^0\gamma^1\gamma^2\gamma^3$, $\sigma_{\mu\nu}=(\gamma_{\mu}\gamma_{\nu}-\gamma_{\nu}\gamma_{\mu})/2$.

In the momentum representation it is convenient to represent the renormalized polarization operator in an external electromagnetic field in the form  (see \cite{virit} and References there)
\bb
\Pi^{\mu \nu}_r(p,k;A) = \Pi^{\mu \nu}_r(p,k;A)-\Pi^{\mu \nu}(p,k;A=0)+\Pi^{\mu \nu}_{0r}(p,k),
\label{porF}
\ee
where $p\equiv p^{\mu}$ is the photon four-momentum,
\bb
\Pi^{\mu \nu}_{0r}(p,k)=\frac{e^2}{2\pi}\delta(p-k)[k^{\mu}p^{\nu}-g^{\mu \nu}p\cdot k]\int_0^1du(1-u^2)\ln\left[1-\frac{p^2}{4m^2}(1-u^2)\right]
\label{pov}
\ee
is the renormalized polarization operator in vacuum and $g^{\mu\nu}$ is the Minkowski tensor $g^{11}=g^{22}=g^{33}-g^{00}=-1$.

Main properties and tensor structure of PO can be obtained from the requirements of
relativistic and gauge invariance and also from the symmetry of external field.
In the momentum representation PO in constant crossed electromagnetic field must be diagonal with respect
to the four-momentum of photon and  can be expressed via three scalar functions $f_1, f_2, f_3$
depending on invariant variables
\bb
\frac{p^2}{m^2}, \quad b^2=-\frac{e^2(F_{\mu\nu}p^{\nu})^2}{m^6},
\label{inv}
\ee
where  $F^{\mu\nu}$ is the tensor of crossed electromagnetic field.
Note that in the special coordinate system with $x^3\equiv z$ axis, aligned with the vector $[{\bf E \times B}]$,
$b^2=e^2p_-^2{\bf E}^2/m^6,\quad p_-=p^0-p^3$.

Finally, PO can be expanded in its eigenvectors as follows
\bb
\Pi^{\mu \nu}_r(p,k;A) = \frac{l^{\mu}_1l^{\nu}_1}{(l_1)^2}(f_1+f_3)+\frac{l^{\mu}_2l^{\nu}_2}{(l_2)^2}(f_2+f_3)+ \frac{l^{\mu}_3l^{\nu}_3}{(l_3)^2}f_3.
\label{por1}
\ee
Here vectors $l^{\mu}_i, i=1, 2, 3$ is determined by
\bb
l^{\mu}_1= F^{\mu\nu}p_{\nu}, \quad l^{\mu}_2=F^{*\mu\nu}p_{\nu}, \quad  l^{\mu}_3=\frac{p^2}{(l_1)^2}F^{\mu\nu}F_{\nu\rho}p^{\rho}+p^{\mu},
\label{eivect}
\ee
and  invariant scalar functions $f_1, f_2, f_3$  can be written as
\bb
f_{1,2}=\frac{e^2m^2b^2}{\pi}\int\limits_0^{\infty}xdx
\int\limits_0^1 du \frac{1-u^2}{48}[9-u^2\mp 3(1-u^2)] \exp\left[-ix\left(1-\frac{p^2(1-u^2)}{4m^2}+\frac{b^2}{48}x^2(1-u^2)^2\right)\right], \nonumber\\
f_3=-\frac{e^2p^2}{2\pi}\int\limits_0^{\infty}\frac{dx}{x}\int\limits_0^1 du (1-u^2) \left[e^{\left(-ix\left(1-\frac{p^2}{4m^2}(1-u^2)+\frac{b^2}{48}x^2(1-u^2)^2\right)\right)}-e^{-ix}\right].\phantom{mmmmmm}
\label{f1,2,3}
\ee
 On the mass shell ($p^2=0$), PO (\ref{por1}) is related to elastic scattering amplitude  of  photon as follows
\bb
A_{1,2}=\frac{e^{*'}_{\mu} \Pi^{\mu \nu}_r(p;A)e_{\nu}}{2p_0},
\label{amp1}
\ee
where $e_{\mu}, e'_{\nu}$ are the polarization vectors of photon in initial and final states and $A_i$ is ESA of photon propagating opposite the vector $[{\bf E \times B}]$ and polarized along ${\bf E}$ ($A_1$) or perpendicularly to ${\bf E}$ ($A_2$).

We need to find  $A_{1,2}$ at $m=0$. In order to integrate $A_{1,2}$ we use the Mellin transform in
parameter $z=\sqrt{48b^2}$:
\bb
A_{1,2}(s)=\int_{0}^{\infty}z^{s-1}A_{1,2}(z)dz.
\label{mel}
\ee
Inverse Mellin transform is written in the form
\bb
A_{1,2}(z)=\frac{1}{2\pi i}\int\limits_{c-i\infty}^{c+i\infty}z^{-s}A_{1,2}(s)ds,\quad c\geq 0.
\label{invmel}
\ee
The right of equation (\ref{invmel}) is  Mellin-Barnes integral \cite{htf} which can be represented via Meijer G-function \cite{beal}
\bb
G^{m n}_{k l}\left(x\left|\begin{array}{c}
a_1, \dots a_k \\
b_1, \dots b_l
\end{array}\right.\right) = \frac{1}{2\pi i}\int_{L} \frac{\prod\limits_{j=1}^{m}\Gamma(b_j-s)\prod\limits_{j=1}^{n}\Gamma(1-a_j+s)}
{\prod\limits_{j=m+1}^{l}\Gamma(1-b_j+s)\prod\limits_{j=n+1}^{k}\Gamma(a_j-s)}x^s ds,
\label{gfun}
\ee
where $L$ is a way separating the poles of gamma function $\Gamma(b_1-s) \dots \Gamma(b_m-s)$ from poles of
$\Gamma(1-a_1+s)\dots)\Gamma(1-a_n+s)$.

Elastic scattering amplitude of photon was obtained in \cite{lrtk} in the form
\bb
A_{1,2}(z)=\frac{e^2m^2}{8p^0 (6\pi)^{3/2}}\left[(17\mp 3)G^{5 1}_{3 5}\left(ze^{i\pi}\left|\begin{array}{c}
0,1/4,3/4  \\
0,0,1/2,1/3,-1/3
\end{array}\right.\right) + \right. \nonumber\\
\left.+(1\mp 3)G^{5 1}_{3 5}\left(ze^{i\pi}\left|\begin{array}{c}
0,-1/4,5/4  \\
0,0,1/2,1/3,-1/3
\end{array}\right.\right)\right].
\label{finamp}
\ee
where $z\equiv 16b^2/3$.

In order to find ESA of photon at $m=0$ we must integrate Eq. (\ref{finamp}) over $s$ closing the contour of integration
at the right. Integral (\ref{finamp}) taken along  such a closed contour reduces to the sum of residues in the poles  lying in the right semi-plane and as a result we find
\bb
A_{1,2}= \frac{e^2}{2^{2/3}3^{1/3}p^0}\left(\frac{(5\mp 1)\Gamma(5/6)}{14\Gamma(7/6)}(3^{-1/2}-i)(e p_-|{\bf E}|)^{2/3}\right).
\label{finres}
\ee
Elastic scattering amplitude of photon defines its "mass" squared that can be related to
the complex index of refraction of some "effective material medium".
The imaginary part of ESA of photon in an external electromagnetic field defines the total probability of
the creation of charged fermion pair  $w$ as follows $w_{1,2}=-2{\rm Im}A_{1,2}$. Thus, the total probability of
the creation of charged massless fermion pair in crossed constant electromagnetic field is
\bb
w_{1,2}= \frac{e^2 2^{1/3}}{3^{1/3}p^0}\frac{(5\mp 1)\Gamma(5/6)}{14\Gamma(7/6)}(e p_-|{\bf E}|)^{2/3}.
\label{prob}
\ee
It is worth while noting that formula (\ref{prob}) is exact.

\section{Elastic scattering amplitude of charged massless fermion in a crossed electromagnetic field}

In the coordinate representation, the mass operator of  charged massive fermion in crossed constant electromagnetic field (in $e^2$-order) is given by
\bb
M(x,x'; A)= -ie^2\gamma^{\mu}S^c(x,x'; A)\gamma^{\nu}S_{\mu\nu}(x-x'),
\label{MO1}
\ee
where $S_{\mu\nu}(x-x')$ is the photon propagation function.

In the momentum representation the renormalized mass operator in considered electromagnetic field is diagonal with respect
to the fermion four-momentum and  can be written as
\bb
M_r(p,A) = M(p,A)-M(p,A=0)+M_r^0(p),
\label{MOr}
\ee
where $p\equiv p^{\mu}$ is the fermion four-momentum and $M_r^0(p)$ is the renormalized mass operator in vacuum.

On the mass shell ($p^2=m^2$), the matrix elements of mass operator (\ref{MOr}) is related to the elastic scattering amplitude  of  charged fermion in crossed electromagnetic field which depends on invariant dynamical variable
\bb
\xi^2=-\frac{e^2(F_{\mu\nu}p^{\nu})^2}{m^6}.
\label{dyn}
\ee
In the coordinate system with $x^3\equiv z$ axis, aligned with the vector $[{\bf E \times B}]$  $\xi^2=e^2p_-^2{\bf E}^2/m^6,\quad p_-=p^0-p^3$. ESA of fermion also depends from the fermion spin  in initial and final states.

Like ESA of photon, elastic scattering amplitude of fermion in crossed electromagnetic field can be written via Meijer G-functions. Since ESA of fermion does not depend on fermion spin in the limit $m=0$, we give ESA of massive fermion summed and averaged in the fermion spin respectively  in final and initial state
\bb
A_e(z_e)=\frac{e^2m^2}{4p^0\sqrt{3} \pi)^3}\left[\frac{5}{6}G^{5 3}_{3 5}\left(z_ee^{i\pi}\left|\begin{array}{c}
0,0,1/2  \\
0,0,-1/3,1/3,1/2
\end{array}\right.\right) + \right. \nonumber\\
\left.-G^{5 3}_{3 5}\left(z_ee^{i\pi}\left|\begin{array}{c}
0,0,-1/2  \\
0,-1/3,1/3,1/2,1
\end{array}\right.\right)\right],
\label{finampe}
\ee
where $z_e\equiv (3\xi)^{-2}$.

ESA of fermion at $m=0$ is obtained by integrating Eq. (\ref{finampe}) over $s$ closing the contour of integration
at the right. Integral (\ref{finamp}) taken along  such a closed contour reduces to the sum of residues in the poles  lying in the right semi-plane and we obtain
\bb
A_e=\frac{e^2 7 3^{2/3}}{27p^0\sqrt{3}}\Gamma(2/3)(1-i\sqrt{3})(e p_-|{\bf E}|)^{2/3}.
\label{finrese}
\ee

The imaginary part of ESA of fermion  defines the total probability of
 photon emission  $w_e$ by fermion as follows $w_e=-2{\rm Im}A_e$. We give the expression for $w_e$ with restored
 the units $c$ and $\hbar$
\bb
w_e= \frac{14e^2 3^{2/3}}{27c\hbar^2 E\sqrt{3}}\Gamma(2/3)\sqrt{3}(ec\hbar E_-|{\bf E}|)^{2/3},
\label{probe}
\ee
where $E$ is fermion energy, $E_-=E-c{\bf p\cdot n}$, ${\bf n}=[{\bf E \times B}]/|[{\bf E \times B}]|$.
This expression, non-perturbative in $|{\bf E}|$, is valid for any $|{\bf E}|$, provided $E_->0$.

\section{Resume}

One-loop radiative corrections to the photon propagation and charged massless fermion motion in external constant crossed electromagnetic field are obtained. We find the massless limits of elastic scattering amplitudes of photon and charged massless fermion in crossed electromagnetic field calculated  earlier in the one-loop approximation of massive quantum electrodynamics. We assume that  the imaginary part of EAS of photon is related to the total probability of charged massless fermion pair creation and show that hypothetical charged massless fermion pair can be created by a photon in  electromagnetic field. We also study the photon emission by a charged massless fermion in the crossed electromagnetic field and obtain the total probability of photon emission calculating elastic scattering amplitude of charged massive fermion in  crossed electromagnetic field in the massless limit.

\end{document}